# IMPLEMENTATION AND EXPERIENCE WITH LUMINOSITY LEVELLING WITH OFFSET BEAM

F. Follin, D. Jacquet, CERN, Geneva, Switzerland

*Abstract*

The practice of luminosity levelling with an offset beam has been used as a routine operation in the LHC since 2011. This paper will describe how it has been implemented and what has been the operational experience with the system.

## INTRODUCTION

The LHC has many experiments, all with different objectives and different luminosity needs. CMS and ATLAS are working with a high-luminosity beam ($8 \times 10^{33}$ cm$^{-2}$ s$^{-1}$ in 2012), whereas LHCb's optimal luminosity is $4 \times 10^{32}$ cm$^{-2}$ s$^{-1}$ and ALICE's working point is around $10^{30}$ cm$^{-2}$ s$^{-1}$. Limiting the luminosity and the pile-up in LHCb and ALICE is essential for data quality [1]. High luminosity could also be responsible for premature ageing of their detector. For ALICE, detectors could also be damaged by high luminosity peak.

The $\beta^*$ value and the number of collisions at each interaction point are optimized for the experiments' needs, but this is not enough to cover for the large range of luminosity needs. In addition, the integrated luminosity for these experiments has to be maximized and the peak luminosity kept under control at the same time. The solution is luminosity levelling.

Among all the possible levelling techniques [2], the levelling by transverse beam offset has been chosen for its flexibility and large range, and the relative simplicity of its implementation. In 2011, the levelling was done manually by the operators before being automated from 2012.

## IMPLEMENTATION IN THE LHC[3]

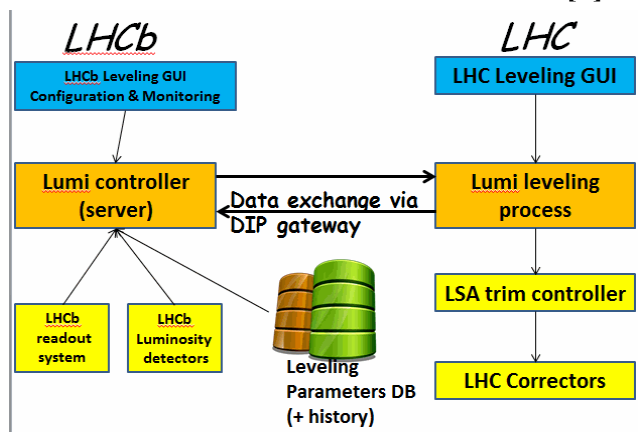

Figure 1: Levelling implementation in LHCb and in LHC.

Control of the levelling has to be implemented, both on the experiment's side and on the LHC side. For example, in the LHCb, a server is responsible for luminosity control. From the LHCb luminosity detectors and LHCb readout system the server publishes the current luminosity to the LHC levelling process, together with the other levelling parameters that are stored in a database with a complete history. A user interface in the LHCb control room has been implemented for levelling monitoring and configuration. The levelling is then completely controlled by the experiments to fulfil their needs (Fig. 1).

### Data Exchange Via DIP Gateway

DIP is the data interexchange protocol that is used for all communication between the LHC and the experiments. For the levelling it is used by the experiments to publish the levelling parameters that have to be used by the levelling process. These parameters are the following:

- **Target luminosity** [$10^{30}$ cm$^{-2}$ s$^{-1}$]: LHCb proton typical target = 400 [$10^{30}$ cm$^{-2}$ s$^{-1}$]; ALICE pPb typical target = 100 [$10^{27}$ cm$^{-2}$ s$^{-1}$]
- **Instant luminosity** [$10^{30}$ cm$^{-2}$ s$^{-1}$]
- **Leveling step size** [σ] (optional): LHCb step size during ramp lumi = 0.2 σ (10.3 μm); LHCb step size when stable lumi= 0.03 σ (1.5 μm)
- **Data quality** (if bad quality, levelling not permitted)
- **Levelling request** (if no request. levelling not permitted)

The LHC levelling application publishes via DIP the levelling status and the status of the crossing plane optimization to LHCb: as long as the crossing is not optimized, LHCb doesn't allow any luminosity levelling.

In the LHC, the luminosity levelling control is part of a more general application that includes also the automation of luminosity scans. The user interface displays the parameters published by the experiments and the instantaneous luminosity and target luminosity evolution with time, as shown in Figs. 2 and 3. The operation team can choose to use either the parameters published by the experiment or the parameters set locally via the user interface.

The levelling algorithm (Fig. 4) is based on a feedback loop on the instantaneous luminosity. The levelling is started by the LHC operation team via the user interface. The instantaneous luminosity is published by the experiments via DIP, the levelling controller does an averaging over several measurements and checks the stability. If the luminosity is in the range defined by the experiments, the measurement loop continues, otherwise a manual action from the LHC operator is requested to changing the separation between the two beams.

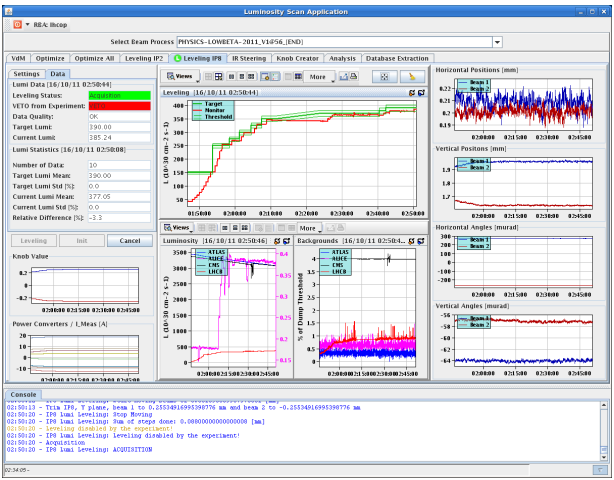

Figure 2: Luminosity scan application.

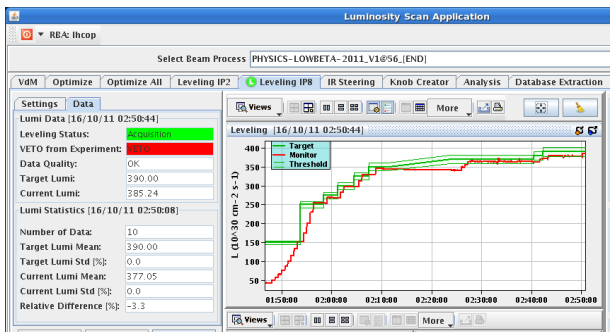

Figure 3: Detail on LHCb levelling control.

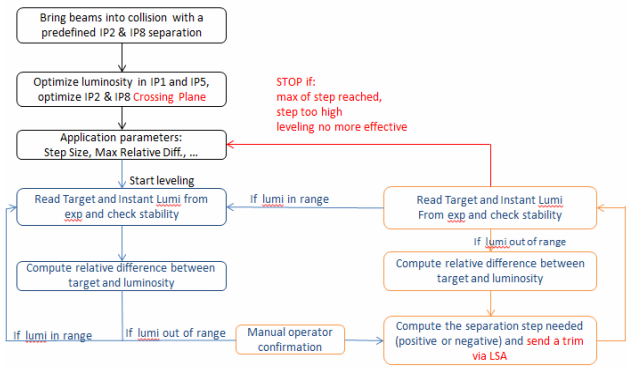

Figure 4: Levelling algorithm.

In current operation, the step size is taken from the experiment's published parameters. After the trim of beam separation, the luminosity reading is checked for stability and whether the value increased or decreased according to the need. If required, the step is undone and inverted. Beams are moved until the luminosity has been pushed within the limits defined by the experiments. When approaching the target, the levelling step is reduced automatically by the algorithm to avoid luminosity overshoot.

The levelling is automatically stopped in the following cases:

- The predefined maximum number of steps has been reached.
- The levelling step is too high
- The levelling is not efficient anymore: beams are in a fully head-on configuration.

*Levelling and LSA*

LSA is the software infrastructure for CERN accelerator control. In LSA database, all the LHC parameters are defined. A hierarchy system links beam parameters to hardware parameters and the rules to computes their values are programmed in the trim package. High level parameters (i.e. tune, beam position at IPs, chromaticity) are called knobs and represent a property of the beams. Their values are change in operation to optimize the beam or change its property and this trim is propagated to the hardware level, i.e. a new current value for a group of magnets.

To change the luminosity, the levelling process computes the step size from sigma to millimetres. It uses the LSA trim package (Fig. 5) and changes the value of knobs that define the beam position in horizontal and vertical plane.

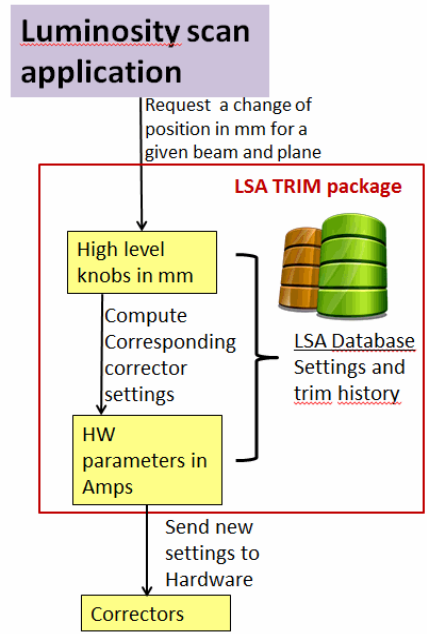

Figure 5: Levelling and LSA trim package.

In LSA, four knobs per IP are defined in units of mm to move each beam in the horizontal or the vertical planes (Fig. 6). Fore correctors are used to control the beam position and angle at the interaction point for a given beam and plane. Each time a new beam position is requested by the levelling, LSA compute the new current in these correctors. Knobs also exist to change the angle in µrad units, but in operation the angle at interaction points is kept to 0. Every settings modification is stored in the LSA database and can be retrieved thanks to the trim history.

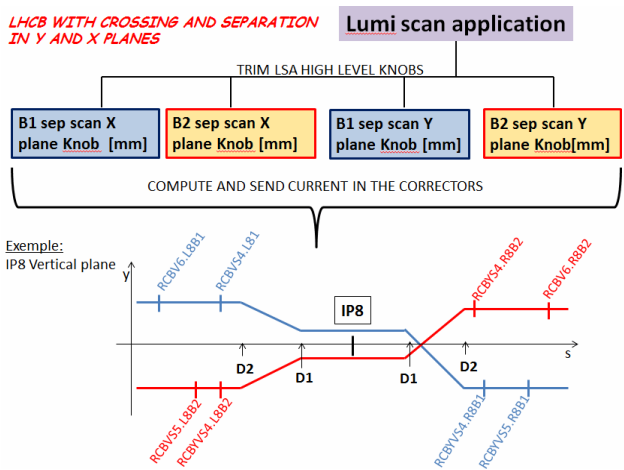

Figure 6: Separation knobs for LHCb.

In 2012, collisions at LHCb were established with a so-called 'tilted' crossing angle to ease the re-setup required at every spectrometer polarity change. The parameter space had to be adapted accordingly, so that higher level knobs were created to move the beams in the crossing and levelling planes. For a given beam, both horizontal and vertical knobs are combined now to move the beam in crossing or levelling plane (Fig. 7).

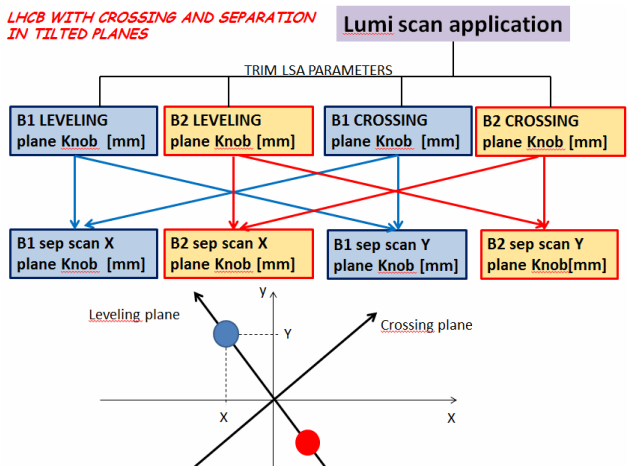

Figure 7: Levelling knobs with LHCb tilted plane.

# OPERATIONAL EXPERIENCE

On 21 April 2011, the luminosity levelling in LHCb was used for the first time in operation. It was followed by the ALICE experiment in May 2011. Thanks to the levelling, the year objective of 1 fb$^{-1}$ integrated luminosity for LHCb was already reached in October 2011, well before the end of the proton run.

In 2012, with the increase of the bunch intensity, the levelling in LHCb was needed for each fill. ALICE chose to run on collisions with satellites, to reduce its luminosity. The levelling was nevertheless needed from time to time depending on the satellite intensity.

The levelling was also prepared and tested for ATLAS and CMS in case the high pileup would become a data quality limitation for these experiments.

In 2013, with the proton–ions run, the levelling was used in ALICE, first during the few days of low luminosity run to keep the luminosity very low and constant. Then at the beginning of each fill to ensure that the luminosity stays beyond the limit requested by ALICE.

*Weakness*

The levelling worked very well with no major issues during two years. Nevertheless, some weakness has been identified:
- The DIP gateway is not always reliable enough and fails sometimes to publish data: this impairs the levelling, as the instantaneous luminosity is not received by the application.
- The luminosity controlled by offset levelling is very sensitive to orbit corrections that are applied regularly during physics to keep the other interaction points at their optimum luminosity. Orbit correction can push the luminosity beyond the limit and in extreme cases trip detectors in LHCb. Even if this is not destructive, this should be avoided, but no preventive mechanism has been implemented on the machine side for now.
- To enable very efficient luminosity control, the experiments have to properly publish the data, e.g. the luminosity target. This was always the case for LHCb, but for ALICE it could have been better managed to gain efficiency.
- For the moment, the algorithm always requires the LHC operator to confirm before starting to move the beams. From time to time this operator response is not immediate and the luminosity continues to go down for several minutes. This could be avoided if the process was fully automated. On the other hand, the operation team needs to check the machine condition before giving the OK for the levelling, for example that no orbit correction is being sent at the same time.

*Observed instabilities*

As already largely discussed in these proceedings [4,5], at the beginning of 2012 run, bunch-by-bunch instabilities were observed (see Figs. 8 and 9). They occurred either in the process of putting beams into collision or once already in stable beams. These instabilities only affected bunches colliding exclusively in IP8.

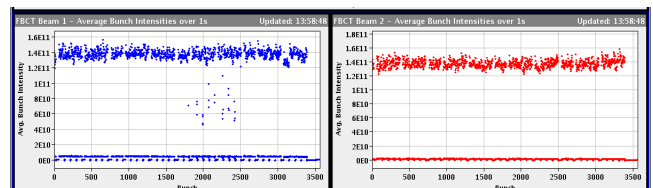

Figure 8: Single bunch instabilities at the beginning of collisions; we see on the bunch-by-bunch intensity plot the intensity drop on Beam 1 bunches colliding in IP8 only.

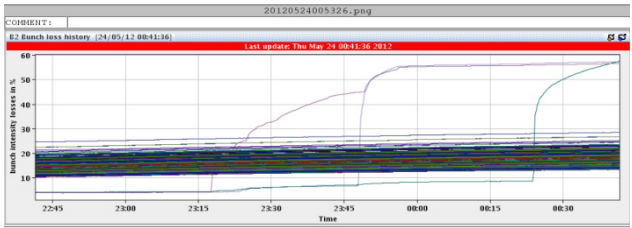

Figure 9: Bunch-by-bunch losses for Beam 2. Beam losses were observed for the three bunches colliding exclusively in IP8 due to instabilities during stable beams.

The first obvious cure that was put in place in operation was to use filling schemes without private bunches for LHCb. Bunches colliding in IP8 are also colliding in IP1 and IP5 and are stabilized by head-on landau damping.

Until the 2012 run, all IPs were put into collision at the same time. To reduce the instabilities observed during this process, this operation was split into two parts. First IP1 and IP5 are put in collision to stabilize the beam as soon as possible. Then the process to tilt the IP8 crossing plane and reduce beam separation in IP8 is played. These solutions have considerably reduced the instabilities.

*Example: LHCb Levelling Proton Run*

The levelling in IP8 is started after IP1 and IP5 are optimized. IP8 has to be optimized in the crossing plane before LHCb gives the permit to start levelling (Fig.10).

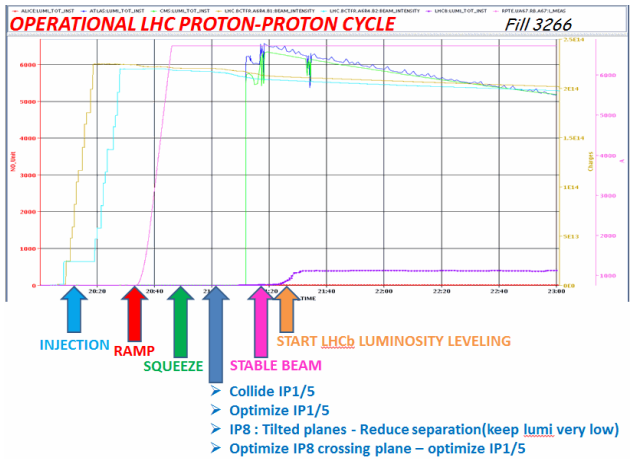

Figure 10: Complete LHC cycle from injection to stable beams.

Figure 11 shows that the initial luminosity of LHCb is very low (less than 10% of the target luminosity). Once the levelling is started, LHCb publishes an intermediate target and request big steps of 0.2 σ in order to reach quickly the target. This time of progressive luminosity increase also allows the conditioning of some detectors. One can also observe that when approaching the target, the application automatically reduces the step size to avoid overshoot.

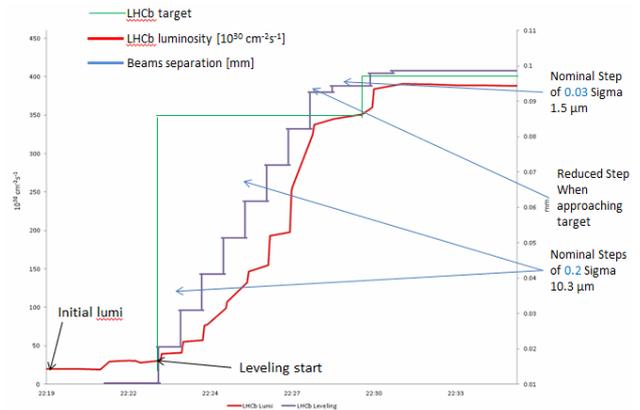

Figure 11: Beginning of levelling process.

After the intermediate target, LHCb publishes the final target that will be used for the rest of the fill. The requested step size is then 0.03 σ to guarantee a maximum stability of the luminosity (Fig. 12).

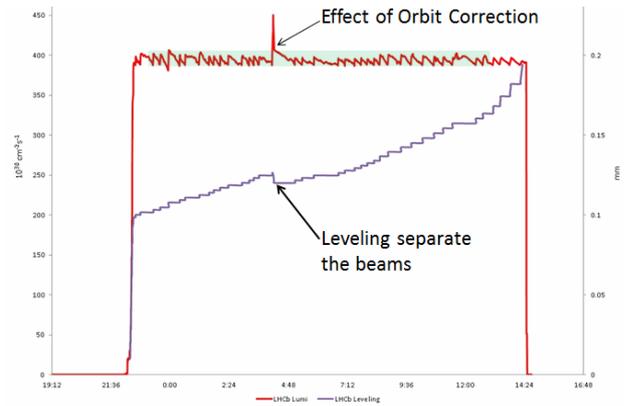

Figure 12: Effect of an orbit correction on the luminosity of LHCb.

*Example: ALICE Levelling Proton–ion Run*

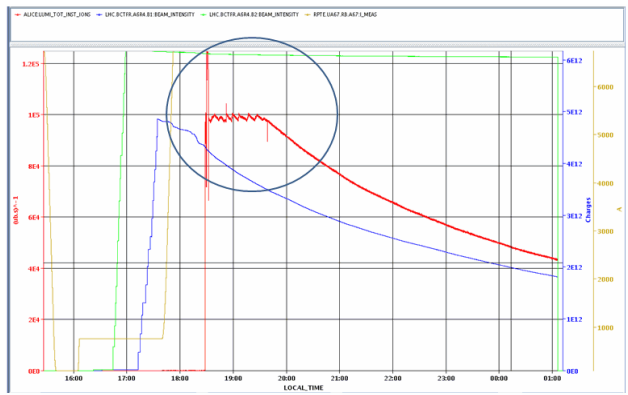

Figure 13: A proton–ion fill and ALICE's luminosity evolution.

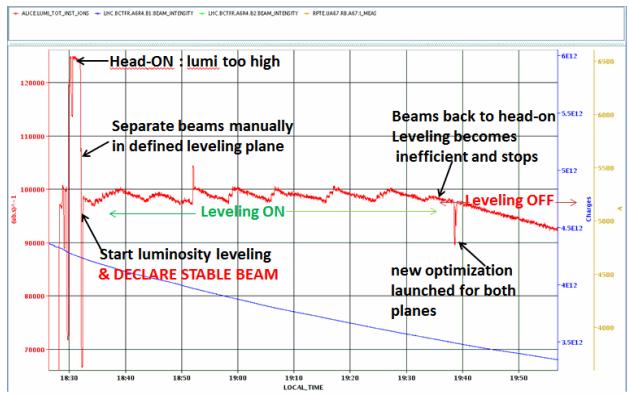

Figure 14: Zoom on ALICE luminosity evolution at the beginning of collisions.

In the example shown in Figs. 13 and 14, during proton–ions physics, ALICE arrives in collision head on with a luminosity higher than the maximum limit. The beams are manually separated until the luminosity is below the target. Then Stable Beams is declared and ALICE levelling started. Luminosity is maintained at the target by the levelling until the beams are back to a head-on configuration, at which time the levelling stops because the luminosity cannot be increased anymore by transversely displacing the beams. The operation team launches a new optimization of IP2 in both planes. The luminosity follows then its natural decay for the rest of the fill.

## CONCLUSION

Luminosity levelling with offset beam has been part of the routine operation since 2011. It allows maximization of the integrated luminosity while keeping the peak luminosity and pileup at the optimum value for the detectors performances. Thanks to the levelling, more than 2 $fb^{-1}$ of exploitable data has been delivered to LHCb in 2012. With 2012 operational conditions, the beam–beam instabilities were under control if using filling schemes with no private bunches for LHCb to ensure head-on landau damping.